\documentclass[12pt,letter]{article}

\usepackage{amsmath, amsthm, amssymb}
\usepackage{amssymb,epsfig}

\def\qed{\hskip 3pt \hbox{\vrule width4pt depth2pt height6pt}}

\newtheorem{Lemma}{Lemma}

\newtheorem{Theorem}[Lemma]{Theorem}
\newtheorem{Proposition}[Lemma]{Proposition}
\newtheorem{Corollary}[Lemma]{Corollary}

\newcommand{\Aut}{\mathop{\mathrm{Aut}}\nolimits}
\newcommand{\Sym}{\mathop{\mathrm{Sym}}\nolimits}

\newcommand{\Cay}{\mathop{\mathrm{Cay}}\nolimits}

\begin{document}

\title{Automorphisms of Cayley graphs generated by transposition sets}
\author{Ashwin Ganesan%
  \thanks{Department of Mathematics, Amrita School of Engineering, Amrita University, Amritanagar, Coimbatore - 641~112, Tamilnadu, India.
 Email: \texttt{ashwin.ganesan@gmail.com, g\_ashwin@cb.amrita.edu}. }}
\date{}

\maketitle


\begin{abstract}
Let $S$ be a set of transpositions such that the girth of the transposition graph of $S$ is at least 5.  It is shown that the automorphism group of the Cayley graph of the permutation group $H$ generated by $S$ is the semidirect product $R(H) \rtimes \Aut(H,S)$, where $R(H)$ is the right regular representation of $H$ and $\Aut(H,S)$ is the set of automorphisms of $H$ that fixes $S$ setwise.  Furthermore, if the connected components of the transposition graph of $S$ are isomorphic to each other, then $\Aut(H,S)$ is isomorphic to the automorphism group of the line graph of the transposition graph of $S$.  This result is a common generalization of previous results by Feng, Ganesan, Harary, Mirafzal, and Zhang and Huang.  
As another special case, we obtain the automorphism group of the extended cube graph that was proposed as a topology for interconnection networks.

\end{abstract}

\bigskip
\noindent\textbf{Keywords} --- Automorphisms of graphs; transpositions; Cayley graphs; extended cube graph.
\\ \noindent \textbf{2010 Mathematics Subject Classification:} 05C25, 05E99.
\bigskip


\section{Introduction}

Let $\Gamma=(V,E)$ be a simple, undirected graph.  An automorphism of $\Gamma$ is a permutation of the vertex set that preserves adjacency, i.e. $\pi \in \Sym(V)$ is an automorphism of $\Gamma$ if $\{u,v\} \in E$ iff $\{u^\pi,v^\pi\} \in E$. The set of automorphisms of the graph $\Gamma$ forms a permutation group, which we denote by $\Aut(\Gamma)$.  

Let $H$ be a group and $S \subseteq H$.  The Cayley digraph of the group $H$ with respect to $S$, denoted by $\Cay(H,S)$, is the digraph with vertex set $H$ and arc set $\{(h,sh): h \in H, s \in S\}$.  When $S$ is closed under inverses, $(g,h)$ is an arc of the digraph $\Cay(H,S)$ if and only if $(h,g)$ is also an arc.  In this case, we identify the two arcs $(g,h)$ and $(h,g)$ with the undirected edge $\{h,g\}$. When the identity element $e$ of $H$ is not in $S$, the Cayley graph $\Cay(H,S)$ has no self-loops.  Thus, when $e \notin S = S^{-1}$, the Cayley digraph $\Cay(H,S)$ can be considered to be a simple, undirected graph.  A Cayley graph $\Cay(H,S)$ is connected if and only if $S$ generates $H$. 

The automorphism group of the Cayley graph $\Cay(H,S)$, denoted by $G:=\Aut(\Cay(H,S))$, contains the right regular representation $R(H)$ as a subgroup; hence all Cayley graphs are vertex-transitive \cite{Biggs:1993}. 
The set of automorphisms of $H$ that fixes $S$ setwise, denoted $\Aut(H,S)$, is a subgroup of the stabilizer $G_e$ in $G$ of $e$ (\cite{Biggs:1993}).  Thus, $R(H) \Aut(H,S) \le G$.  If equality holds, then $\Aut(\Cay(H,S))$ is the semidirect product $R(H) \rtimes \Aut(H,S)$.  In this case, we call the Cayley graph $\Cay(H,S)$ \emph{normal} since $R(H)$ is a normal subgroup of automorphism group of the Cayley graph.  Thus, normal Cayley graphs 
are those 
that have the smallest possible full automorphism group $R(H) \rtimes \Aut(H,S)$.  The automorphism group $G$ of a Cayley graph $\Cay(H,S)$ is equal to the semidirect product $R(H) \rtimes \Aut(H,S)$ if and only if $G_e \subseteq \Aut(H)$ (cf. \cite{Xu:1998}, \cite{Godsil:1981}).  An open problem in the literature is to determine which Cayley graphs are normal.  

In the present paper, we consider the generator set $S$ to be a set of transpositions.  Given a set of transpositions $S$, the transposition graph of $S$ is defined to be the graph whose vertex set is the support of $S$, and two vertices $i$ and $j$ are adjacent in this graph whenever $(i,j) \in S$. For eg, if $S=\{(1,2),(2,3),(1,3)\}$, then the transposition graph of $S$ is a triangle on vertex set $\{1,2,3\}$, and $H:=\langle S \rangle = S_3$ and $\Cay(H,S) \cong K_{3,3}$. In this case, $S$ is not a minimal generating set for $H$.  A generating set $S$ for $H=\langle S \rangle$ is minimal if and only if the transposition graph of $S$ is a tree \cite{Godsil:Royle:2001}.  The Cayley graph of the permutation group generated by a single transposition is trivially understood, so we shall assume henceforth that $S$ is a set of transpositions that contains at least two elements.  

We recall the results from the literature on the automorphism group of Cayley graphs generated by transposition sets.  Let $S$ be a set of transpositions generating the symmetric group $S_n$; in other words, suppose the transposition graph of $S$ is connected and has vertex set $\{1,\ldots,n\}$. Let $G$ denote the automorphism group of the Cayley graph $\Cay(S_n,S)$.  When the transposition graph is an asymmetric tree, Godsil and Royle \cite{Godsil:Royle:2001} showed that $G \cong S_n$.  Zhang and Huang studied the case where the transposition graph is a path graph (\cite{Zhang:Huang:2005}) and where the transposition graph is a star (\cite{Huang:Zhang:STn:submitted}).  Feng \cite{Feng:2006} showed that $\Aut(S_n,S)$ is isomorphic to the automorphism group of the transposition graph of $S$ when $n \ge 3$; this holds even if the transposition graph of $S$ contains cycles or is not connected.  In the special case when the transposition graph of $S$ is a tree, Feng \cite{Feng:2006} also showed that $G \cong R(S_
n) \rtimes \Aut(S_n,S)$.   Ganesan \cite{Ganesan:2012} showed that if the girth of the 
transposition graph is at least 5, then $G \cong R(S_n) \rtimes \Aut(S_n,S)$.  

For a graph $\Gamma$, let $G_v$ denote the set of automorphisms of $\Gamma$ that fix the vertex $v$, and let $L_v$ denote the set of automorphisms of $\Gamma$ that fix the vertex $v$ and each of its neighbors.  Then, $L_v \trianglelefteq G_v$ \cite{Biggs:1993}. Ganesan \cite{Ganesan:2012} showed that if the transposition graph of $S$ is connected and has girth at least 5, then the stabilizer $L_v$ is trivial and $\Cay(S_n,S)$ is normal, whereas if the transposition graph is a 4-cycle, then $L_v$ is isomorphic to the Klein 4-group $\mathbb{Z}_2 \times \mathbb{Z}_2$ and hence is nontrivial.  Thus, this Cayley graph $\Cay(S_4,S)$ is not normal. 

In all of the papers mentioned above on Cayley graphs generated by transpositions, the set of transpositions $S$ is assumed to generate $S_n$.  In the present paper, we consider the more general case where $S$ does not necessarily generate $S_n$, i.e. the transposition graph of $S \subseteq S_n$ is not necessarily connected. Let $H \le S_n$ be the permutation group generated by $S$.  Each connected component in $\Cay(S_n,S)$ is isomorphic to $\Cay(H,S)$ and has as its vertex set a right coset of $H$ in $S_n$.  The $r$-dimensional hypercube graph is defined as the Cayley graph of the permutation group $H$ generated by $r$ disjoint transpositions, i.e. the transposition graph of $S$ consists of $r$ disjoint edges $K_2$ and $H \cong \mathbb{Z}_2^r$.  Harary \cite{Harary:2000} determined the automorphism group of the hypercube graph to be exponentiation group, and Mirafzal \cite{Mirafzal:2011} used a different method to show that the automorphism group of the hypercube graph is $\mathbb{Z}_2^r \rtimes S_r$.  

The objective of the present investigation is to seek a common generalization of the results mentioned above on the automorphism group of Cayley graphs generated by transpositions to the case where the transposition graph of $S$ may contain more than one connected component (such as in the case of the hypercube graph mentioned above), but each connected component might be more general than just being a single edge $K_2$.  For example, each component of the transposition graph of $S$ could be a path graph of arbitrary length, in which case the corresponding Cayley graph is called the extended cube graph.  This graph was proposed in \cite[p. 384]{Lakshmivarahan:etal:1993} as a topology for interconnection networks.  A special case of the main theorem in the present paper gives the automorphism group of this extended cube graph.

When choosing a graph as the topology of an interconnection network, it is desired that the graph have a high degree of symmetry.  To that end, the symmetry properties of topologies proposed for interconnection networks have been extensively studied; for example, the automorphism group of the following topologies of interconnection networks have been studied: the bubble-sort graph \cite{Zhang:Huang:2005}, the star graph \cite{Huang:Zhang:STn:submitted}, the modified bubble sort graph \cite{Zhang:Huang:2005}, the derangement graph \cite{Deng:Zhang:2011}, the folded hypercube \cite{Mirafzal:2011}, and the alternating group graph \cite{Zhou:2011}.     Besides the automorphism group, other properties of Cayley graphs generated by transposition sets have also been investigated; see for example, the papers by Cheng, Lipt\'{a}k and their coauthors \cite{Lin:Cheng:etal:2008}, \cite{Cheng:etal:2012}.

Given a simple, undirected graph $\Gamma=(V,E)$, the line graph of $\Gamma$ is defined as the graph with vertex set $E$, and with two vertices being adjacent in the line graph whenever the corresponding two edges share a common endpoint in $\Gamma$. 

In Section~\ref{sec:main:results} we state the main result of this paper and mention some corollaries and special cases.  Section~\ref{sec:proof:of:main:results} contains the proof of the main theorem. 

\section{Main results}
\label{sec:main:results}

The main result of this paper is the following theorem:

\begin{Theorem} \label{main:theorem}
Let $S$ be a set of transpositions such that the girth of the transposition graph of $S$ is at least 5.  Then, the automorphism group of the Cayley graph of the permutation group $H$ generated by $S$ is the semidirect product $R(H) \rtimes \Aut(H,S)$, where $R(H)$ is the right regular representation of $H$ and $\Aut(H,S)$ is the set of automorphisms of $H$ that fixes $S$ setwise.  Furthermore, if the connected components of the transposition graph of $S$ are isomorphic to each other,  then $\Aut(H,S)$ is isomorphic to the automorphism group of the line graph of the transposition graph of $S$.
\end{Theorem}

We now mention some corollaries and special cases of the theorem.  For convenience, we let $T=T(S)$ denote the transposition graph of $S$, and suppose that $T$ consists of $r$ disjoint copies of some connected graph $\Gamma$ whose girth is at least 5. 

First consider the case where the transposition graph of $S$ is a tree on three or more vertices. The permutation group $H$ is the symmetric group $S_n$ for some $n \ge 3$. In this case $r=1$ since the transposition graph consists of a single connected component. By Theorem~\ref{main:theorem}, the automorphism group of $\Cay(S_n,S)$ is $R(S_n) \rtimes \Aut(S_n,S)$, which is the result given in Feng \cite{Feng:2006}:  

\begin{Corollary} (\cite{Feng:2006})
Let $S$ be a set of transpositions such that the transposition graph of $S$ is a tree on $n$ vertices, where $n \ge 3$.  Then $\Aut(\Cay(S_n,S)) = R(S_n) \rtimes \Aut(S_n,S)$. 
\end{Corollary}

Ganesan \cite{Ganesan:2012} strengthened this result by showing that a sufficient condition for normality of $\Cay(S_n,S)$ is that the connected transposition graph of $S$ have girth at least 5: 

\begin{Corollary} (\cite{Ganesan:2012})
Let $S$ be a set of transpositions such that the transposition graph of $S$ is a connected graph on $n$ vertices and has girth at least 5, where $n \ge 3$.  Then $\Aut(\Cay(S_n,S)) =R(S_n) \rtimes \Aut(S_n,S)$.  
\end{Corollary}

When the transposition graph of $S$ is the $n$-cycle graph, the Cayley graph $\Cay(S_n,S)$ is called the modified bubble sort graph.  Zhang and Huang \cite{Zhang:Huang:2005} showed that the automorphism group of this graph is $D_{2n}S_n$.  Actually, it is shown in Ganesan \cite{Ganesan:2012} that this result is true if and only if $n \ge 5$ (the 4-cycle transposition graph causes a strictly larger automorphism group).  By Theorem~\ref{main:theorem}, when $n \ge 5$, the automorphism group of $\Cay(S_n,S)$ is $R(S_n) \rtimes Aut(S_n,S)$, where $\Aut(S_n,S) \cong D_{2n}$:

\begin{Corollary} (\cite{Ganesan:2012})
Let $S$ be a set of transpositions such that the transposition graph of $S$ is an $n$-cycle graph, $n \ge 5$.  Then $\Aut(\Cay(S_n,S)) \cong R(S_n) \rtimes D_{2n}$.
\end{Corollary}

When the transposition graph of $S$ consists of $r$ independent edges (i.e. $T=rK_2)$, the line graph of the transposition graph of $S$ is $\overline{K}_r$.  The permutation group $H:=\langle S \rangle$ is isomorphic to $\mathbb{Z}_2^r$.  By Theorem~\ref{main:theorem}, it follows that the automorphism group of the hypercube is $\mathbb{Z}_2^r \rtimes S_r$, which is obtained in Harary \cite{Harary:2000} and also given explicitly in Mirafzal \cite[p. 6]{Mirafzal:2011}:

\begin{Corollary} (\cite{Harary:2000}, \cite{Mirafzal:2011})
 The automorphism group of the $r$-dimensional hypercube graph is isomorphic to $\mathbb{Z}_2^r \rtimes S_r$.
\end{Corollary}

Frucht's theorem \cite{Frucht:1949} states that the automorphism group of $r$ disjoint copies of a connected graph $\Gamma$ is the wreath product $S_r[\Aut(\Gamma)]$.  If the transposition graph $T(S)$ on vertex set $\{1,\ldots,n\}$ is not connected, then the permutation group $H$ generated by $S$ is a proper subgroup of $S_n$.  Then the Cayley graph $\Cay(S_n,S)$ consists of $\ell:=|S_n:H|$ connected components.  Each of these connected components is isomorphic to $\Cay(H,S)$ and has as its vertex set a right coset of $H$ in $S_n$.  Thus, the automorphism group of $\Cay(S_n,S)$ can be obtained from the automorphism group of $\Cay(H,S)$:  

\begin{Corollary}
Let $S$ be a set of transpositions in $S_n$ generating $H$ such that the girth of the transposition graph of $S$ is at least 5.  Then $\Aut(\Cay(S_n,S))$ is isomorphic to the wreath product $S_\ell[\Aut(\Cay(H,S))] = S_\ell[R(H) \rtimes \Aut(H,S)]$, where $\ell$ is the index of $H$ in $S_n$. 
\end{Corollary}

Suppose the transposition graph $T=T(S)$ consists of $r$ copies of $P_3$, the path graph on $3$ vertices.  Then $\Cay(H,S)$ is called the extended cube graph of dimension $r$, which was proposed in \cite[p. 384]{Lakshmivarahan:etal:1993} as a topology of interconnection networks.  The line graph of $T$ is $rP_2$, which by Frucht's theorem has automorphism group $S_r[S_2]$. Furthermore, the permutation group $H$ generated by $S$ is isomorphic to the direct product of $r$ copies of $S_3$. Thus, the automorphism group of the extended cube graph is isomorphic to $S_3^r \rtimes S_r[S_2]$.  This result can be generalized further to other extensions of the cube graph, for example, from path graphs on 3 vertices to path graphs on 3 or more vertices, or to disjoint copies of other kinds of graphs.  For $k \ge 3$, define the extended cube graph of dimension $(r,k)$ to be the Cayley graph of the permutation group $H$ generated by a set of transpositions $S$ whose transposition graph consists of $r$ disjoint copies of a 
path graph on $k$ vertices.   Then, we have:

\begin{Corollary} The automorphism group of the extended cube graph of dimension $(r,k)$ is isomorphic to $S_k^r \rtimes S_r[S_2]$. 
\end{Corollary}

\section{Proofs of main results}
\label{sec:proof:of:main:results}

Throughout, $S$ denotes a set of transpositions, $H$ denotes the permutation group generated by $S$, and $e$ denotes the identity element in $H$. 

\begin{Theorem} \label{thm:unique46:implies:normality}
Suppose $S$ is a set of transpositions generating $H$ satisfying the following two conditions for all $t,k \in S, t \ne k$: 
\\(i) $tk=kt$ if and only if there is a unique 4-cycle in $\Cay(H,S)$ containing $e,t$ and $k$.
\\(ii) If $tk \ne kt$, then there is a unique 6-cycle in $\Cay(H,S)$ containing $e,t,k$ and a vertex at distance 3 from $e$.
\\Then, $\Aut(\Cay(H,S)) \cong R(H) \rtimes \Aut(H,S)$.
\end{Theorem}

\noindent \emph{Proof}:  Let $G$ denote the automorphism group of the Cayley graph $\Cay(H,S)$.  By Xu \cite[Proposition 1.5]{Xu:1998}, $G = R(H) \rtimes \Aut(H,S)$ if and only if the stabilizer subgroup $G_e \subseteq \Aut(H)$.  Let $\sigma \in G_e$.   It suffices to show that $\sigma$ is an automorphism of $H$. Since $\sigma$ is an automorphism of the Cayley graph $\Cay(H,S)$, $\sigma$ is a bijection from the vertex set $H$ of the graph to itself.  

To prove that the map $\sigma$ from $H$ to $H$ is a homomorphism, we induct on the length of a word in $H$ in terms of the generator elements.  Let $s_1,s_2,\ldots,s_k \in S$.  We prove by induction on $k$ that $(s_1 \cdots s_k)^\sigma = s_1^\sigma \cdots s_k^\sigma$.  This is clearly true for $k=1$.  Since $S$ satisfies conditions (i) and (ii) given in the hypotheses, by the proof given in \cite[p. 71]{Feng:2006}, $(s_1 s_2)^\sigma=s_1^\sigma s_2^\sigma$.  Now suppose that for all $\sigma \in G_e$, $(s_1 \cdots s_{k-1})^\sigma = s_1^\sigma \cdots s_{k-1}^\sigma$.  Define $z:=s_k, y:=(s_k^\sigma)^{-1}$, and let $r_z$ and $r_y$ denote right translation by $z$ and $y$, respectively.    Observe that $(s_1 \cdots s_k)^\sigma = s_1^\sigma \cdots s_k^\sigma$ if and only if $(s_1 \cdots s_{k-1})^{r_z \sigma r_y} = s_1^\sigma \cdots s_{k-1}^\sigma$. Since $e^{r_z \sigma r_y} = (s_k)^{\sigma r_y} = (s_k^\sigma)y=e$, $r_z \sigma r_y \in G_e$ satisfies the inductive hypothesis.  Thus, $(s_1 \cdots s_{k-1})^{r_z \sigma 
r_y} = (s_1)^{r_z \sigma r_y} \cdots s_{k-1}^{r_z \sigma r_y} = s_1^\sigma \cdots s_{k-1}^\sigma$.\qed

A special case of the next lemma is proved in Godsil and Royle \cite[Lemma 3.10.3]{Godsil:Royle:2001}, where it is assumed that $S$ is a set of transpositions such that the transposition graph of $S$ does not contain triangles.  The lemma remains valid even if the transposition graph of $S$ contains triangles: 

\begin{Lemma} \label{lemma:unique4}
Let $S$ be any set of transpositions and let $H$ be the permutation group generated by $S$.  Let $t,k \in S, t \ne k$. Then, $tk=kt$ if and only if there is a unique 4-cycle in $\Cay(H,S)$ containing $e,t$ and $k$.
\end{Lemma}

\noindent \emph{Proof}: Suppose $tk=kt$.  Then $t$ and $k$ have disjoint support.  Let $w$ be a common neighbor of the vertices $t$ and $k$ in the Cayley graph $\Cay(H,S)$.  By definition of the adjacency relation in the Cayley graph, there exist $x,y\in S$ such that $xt=yk=w$. Observe that $xt=yk$ iff $tk=xy$.  But since $k$ and $t$ have disjoint support, $tk=xy$ iff $t=x$ and $k=y$ or $t=y$ and $k=x$.  Thus, $w$ is either the vertex $e$ or the vertex $tk$.  Hence, there exists a unique 4-cycle in $\Cay(H,S)$ containing $e,t$ and $k$, namely the cycle $(e,t,tk=kt,k,e)$.  

To prove the converse, suppose $tk \ne kt$.  Then $t$ and $k$ have overlapping support; without loss of generality, take $t=(1,2)$ and $k=(2,3)$.  We consider two cases, depending on whether $(1,3) \in S$. First suppose $(1,3) \notin S$. Let $w$ be a common neighbor of $t$ and $k$.  So $w=xt=yk$ for some $x,y \in S$.  As before, $xt=yk$ iff $xy=tk=(1,2)(2,3)=(1,3,2)$.  The only ways to decompose $(1,3,2)$ as a product of two transpositions is as $(1,3,2)=(1,2)(2,3)=(3,2)(1,3)=(1,3)(1,2)$.  Since $(1,3) \notin S$, we must have $x=(1,2)$ and $y=(2,3)$, whence $w=e$.  Thus, $t$ and $k$ have only one common neighbor, namely $e$. Therefore, there does not exist any 4-cycle in $\Cay(H,S)$ containing $e,t$ and $k$.

Now suppose $r:=(1,3) \in S$.  Then $S$ contains the three transpositions $t=(1,2),k=(2,3)$ and $r=(1,3)$.  The Cayley graph of the permutation group generated by these transpositions is the complete bipartite graph $K_{3,3}$. Hence $\Cay(H,S)$ contains as a subgraph the complete bipartite graph $K_{3,3}$ with bipartition $\{e,kt,tk\}$ and $\{t,k,r\}$.  There are exactly two  4-cycles in $\Cay(H,S)$ containing $e,t$ and $k$, namely the 4-cycle through the vertex $kt$ and the 4-cycle through the vertex $kt$. Thus, while there exists a 4-cycle in this case, it is not unique.
\qed

If $H'$ is a permutation group containing $H$, then the Cayley graph $\Cay(H',S)$ contains the Cayley graph $\Cay(H,S)$ as one of its connected components.  In fact, the vertex set of each connected component of $\Cay(H',S)$ is one of the right cosets of $H$ in $H'$.  Thus, there is a unique 4-cycle in $\Cay(H',S)$ containing $e,t$ and $k$ if and only if there is a unique 4-cycle in $\Cay(H,S)$ containing $e,t$ and $k$.  

\begin{Theorem} \label{girth:5:implies:unique46}
Let $S$ be a set of transpositions such that the girth of the transposition graph is at least 5.  Then the automorphism group of the Cayley graph of the permutation group $H$ generated by $S$ is the semidirect product $R(H) \rtimes \Aut(H,S)$. 
\end{Theorem}

\noindent \emph{Proof}:  It suffices to prove that if $S$ is a set of transpositions such that the girth of the transposition graph of $S$ is at least 5, then conditions (i) and (ii) of Theorem~\ref{thm:unique46:implies:normality} are satisfied.  Let $t,k \in S, t \ne k$.  By Lemma~\ref{lemma:unique4}, $tk=kt$ if and only if there is a unique 4-cycle in $\Cay(H,S)$ containing $e,t,$ and $k$.  Thus, condition (i) of Theorem~\ref{thm:unique46:implies:normality} is satisfied.  For the special case where the transposition graph of $S$ has girth at least 5 and is a connected graph, it is shown in \cite{Ganesan:2012} that there is a unique 6-cycle in $\Cay(H,S)$ containing $e,t,k$ and a vertex at distance 3 from $e$.  It can be checked that this identical proof goes through even if the transposition graph of $S$ is not connected since this proof in \cite{Ganesan:2012} does not use the assumption of connectivity of the transposition graph.  Thus, condition (ii) of Theorem~\ref{thm:unique46:implies:normality} is 
also satisfied.  Hence, $\Aut(\Cay(H,S)) = R(H) \rtimes \Aut(H,S)$.\qed  

\bigskip

\begin{Proposition} \label{prop:autHS:in:autLT}
Let $S$ be any set of transpositions, and let $H$ be the permutation group generated by $S$.  Then, every automorphism of $H$ that fixes $S$ setwise, when restricted to $S$, is an automorphism of the line graph of the transposition graph of $S$. 
\end{Proposition}

\noindent \emph{Proof}: Let $\pi \in \Aut(H,S)$. Let $t,k \in S, t \ne k$. Since $\pi$ is an automorphism of $H$, it takes $tk$ to $(tk)^\pi = t^\pi k^\pi$.  An automorphism of a group preserves the order of the elements, whence $t$ and $k$ have disjoint support if and only if $t^\pi$ and $k^\pi$ have disjoint support.  Since $\pi$ fixes $S$, $t^\pi, k^\pi \in S$.  Thus, in the transposition graph of $S$, the edges $t$ and $k$ are incident if and only if the edges $t^\pi$ and $k^\pi$ are incident.  In other words, $\pi$ restricted to $S$ is an automorphism of the line graph of the transposition graph of $S$.\qed

\begin{Theorem}
 Let $S$ be a set of transpositions such that the connected components of the transposition graph of $S$ are isomorphic to each other and have girth at least 5.  Let $H$ be the permutation group generated by $S$. Then, $\Aut(H,S)$ is isomorphic to the automorphism group of the line graph of the transposition graph of $S$.
\end{Theorem}

\noindent \emph{Proof}: For simplicity of exposition, we denote the transposition graph of $S$ by $T=T(S)$ and its line graph by $L(T)$. By hypothesis, the transposition graph $T$ consists of $r$ disjoint copies of some connected graph $\Gamma$ for some positive integer $r$, and the girth of $T$, which equals the girth of $\Gamma$, is at least 5.  

If $\Gamma$ has exactly 2 vertices, then $\Gamma$ is the single edge $K_2$.  In this case, the line graph of $T=rK_2$ is $\overline{K}_r$ and has automorphism group $S_r$.  The permutation group $H$ generated by $r$ disjoint transpositions is isomorphic to $\mathbb{Z}_2^k$.  Any automorphism of $H$ is uniquely determined by its image of the $r$ generator elements, and hence $\Aut(H,S)$ has at most $r!$ elements.  Conversely, every permutation of these generators $S$ can be extended to a unique automorphism of $H$.  Thus, $\Aut(H,S) \cong S_r$.  We have shown that that the assertion $\Aut(H,S) \cong \Aut(L(T))$ is true if each connected component $\Gamma$ of the transposition graph $T$ has exactly 2 vertices.

For the rest of the proof, suppose each connected component $\Gamma$ of the transposition graph $T$ has at least 3 vertices. In view of Proposition~\ref{prop:autHS:in:autLT}, we can define the map $\psi: \Aut(H,S) \rightarrow \Aut(L(T))$ that takes $\sigma \in \Aut(H,S)$ to $\sigma$ restricted to $S$.  We denote the image of $\sigma$ by $\sigma|_S$. We show $\psi$ is an isomorphism. It is clear that $\psi$ is a homomorphism. Also, a homomorphism is completely determined by the image of the generators, so that $\sigma \in \Aut(H,S)$ is completely determined by its action on $S$.  Thus, $\psi$ is injective.  We prove next that $\psi$ is onto.  Let $\tau \in \Aut(L(T))$.  Our objective is to show there exists an element $\sigma \in \Aut(H,S)$ such that $\sigma |_S=\tau$. 

Given $\tau \in \Aut(L(T))$, we claim that there exists a unique element $\sigma' \in \Aut(T)$ that induces $\tau$. We first consider the case where $T$ consists of just one connected component.  It is clear that for any two graphs $T, T'$, an isomorphism from $T$ to $T'$ induces an isomorphism from $L(T)$ to $L(T')$.  Whitney \cite{Whitney:1932} showed that if $T,T'$ are connected graphs such that their line graphs $L(T),L(T')$ are isomorphic, then $T,T'$ are also isomorphic unless one is a triangle $K_3$ and the other is a star $K_{1,3}$.  In particular, it can be shown (cf. \cite[p. 73]{Harary:1969}) that if $\Gamma,\Gamma'$ are connected graphs on at least 5 vertices, then for every isomorphism $\phi_1$ from $L(\Gamma)$ to $L(\Gamma')$, there is exactly one isomorphism $\phi$ from $\Gamma$ to $\Gamma'$ that induces $\phi_1$.   Setting $T'$ to equal $T$, it follows that  if the transposition graph $T$ is a connected graph on at least 5 vertices, then for every automorphism $\tau$ 
of $L(T)$ there is a unique automorphism $\sigma'$ of $T$ that induces $\tau$.  Therefore, the automorphism group of $T$ and of its line graph $L(T)$ are isomorphic. If $T$ is a connected graph on 3 or 4 vertices, then by the hypothesis on the girth of $T$, $T$ is a tree.  In this case, the claim can be verified for each tree.  This proves the claim if $T$ consists of just one connected component.  

If $T$ consists of $r$ disjoint copies of some connected graph $\Gamma$, then each connected component of $L(T)$ corresponds to a connected component of $T$.  By the same argument given for the $r=1$ case, it follows that for every automorphism $\tau$ of $L(T)$, there exists an automorphism $\sigma'$ of $T$ that induces $\tau$. 

The vertex set of $T$ is the support of $H$, call it $\{1,\dots,n\}$.  Thus, $H = \langle S \rangle \le S_n$. Given $\tau \in \Aut(L(T))$, we showed there exists an element $\sigma' \in \Aut(T)$ such that $\sigma'$ induces $\tau$.  Given this $\sigma'$, let $c(\sigma')$ denote conjugation by $\sigma'$.  Thus, $c(\sigma') \in \Aut(S_n)$.  Since $\sigma'$ is an automorphism of $T$, it fixes the edge set $S$ of $T$.  Hence, $c(\sigma') \in \Aut(S_n,S)$.  Define $\sigma:= c(\sigma') |_H$.  Then $\sigma$ is an automorphism of $H$ that fixes $S$ setwise.  Also, $\sigma|_S=\tau$.  For example, if $\tau$ takes $\{i,j\}$ to $\{m,\ell\}$, then there exists a $\sigma' \in S_n$ that takes $\{i,j\}$ to $\{\sigma'(i),\sigma'(j)\} = \{m,\ell\}$.  Then $c(\sigma')$ takes $(i,j) \in S$ to $(m,\ell) \in S$.  Thus, $c(\sigma')|_H$ and $\tau$ induce the same permutation of $S$, which implies $\psi$ is onto.  We have shown that $\Aut(H,S) \cong \Aut(L(T))$.\qed

\bibliographystyle{plain}
\bibliography{refsaut}

\end{document}